\documentclass[journal=apchd5,manuscript=article]{achemso}

\usepackage[version=3]{mhchem}
\usepackage{graphicx}% Include figure files
\usepackage{multicol}
\usepackage{setspace}
\usepackage{achemso}
 \setkeys{acs}{usetitle = true}
\title{\textbf{Simultaneous Multi-Harmonic Imaging of Nanoparticles in Tissues for Increased Selectivity}}

\author{Andrii Rogov}
\affiliation[GAP-Biophotonics, Universit\'e de Gen\`eve, 22 Chemin de Pinchat, 1211 Gen\`eve, Switzerland]{GAP-Biophotonics, University of Geneva}
\author{Marie Irondelle}
\affiliation[Cell and Tissue Imaging Facility (PICT-IBiSA), CNRS UMR144 Institut Curie - Paris 75248, France]{CNRS UMR144 Institut Curie, Paris}
\author{Fernanda Ramos-Gomes}
\author{Julia Bode}
\affiliation[Department of Molecular Biology of Neuronal Signals, Max-Planck-Institute of Experimental Medicine, Hermann-Rein-Str. 3, 37075 G\"ottingen, Germany]{ Max-Planck-Institute of Experimental Medicine, G\"ottingen} 
\author{Davide Staedler}
\affiliation[GAP-Biophotonics, Universit\'e de Gen\`eve, 22 Chemin de Pinchat, 1211 Gen\`eve, Switzerland]{GAP-Biophotonics, University of Geneva}
\alsoaffiliation[Laboratory of Synthesis and Natural Products, Institute of Chemical Sciences and Engineering, \'Ecole Polytechnique F\'ed\'erale de Lausanne, Batochime, 1015 Lausanne, Switzerland]{LSNP, EPF Lausanne} 
\author{Sol\`ene Passemard}
\affiliation[Laboratory of Synthesis and Natural Products, Institute of Chemical Sciences and Engineering, \'Ecole Polytechnique F\'ed\'erale de Lausanne, Batochime, 1015 Lausanne, Switzerland]{LSNP, EPF Lausanne} 
\author{S\'ebastien Courvoisier}
\affiliation[GAP-Biophotonics, Universit\'e de Gen\`eve, 22 Chemin de Pinchat, 1211 Gen\`eve, Switzerland]{GAP-Biophotonics, University of Geneva}
\author{Yasuaki Yamamoto}
\affiliation[JEOL SAS Espace Claude Monet - 1,allee de Giverny 78290 CROISSY-SUR-SEINE(France)
]{JEOL SAS} 
\author{Fran\c{c}ois Waharte} 
\affiliation[Cell and Tissue Imaging Facility (PICT-IBiSA), CNRS UMR144 Institut Curie - Paris 75248, France]{CNRS UMR144 Institut Curie, Paris}
\affiliation[Laboratory of Synthesis and Natural Products, Institute of Chemical Sciences and Engineering, \'Ecole Polytechnique F\'ed\'erale de Lausanne, Batochime, 1015 Lausanne, Switzerland]{Laboratory of Synthesis and Natural Products, EPFL} 
\author{Daniel Ciepielewski}
\affiliation[Nikon AG - Instruments,  Im Hanselmaa 10, 8132 Egg / ZH, Switzerland]{Nikon AG - Instruments} 
\author{Philippe Rideau}
\affiliation[Nikon France \& BeLux - Division Instruments 191, rue du march\'e Rollay - 94504 Champigny sur Marne, France]{Nikon France \& BeLux - Division Instruments}
\author{Sandrine Gerber-Lemaire}\affiliation[Laboratory of Synthesis and Natural Products, Institute of Chemical Sciences and Engineering, \'Ecole Polytechnique F\'ed\'erale de Lausanne, Batochime, 1015 Lausanne, Switzerland]{LSNP, EPF Lausanne} 

\author{Frauke Alves}
\affiliation[Department of Molecular Biology of Neuronal Signals, Max-Planck-Institute of Experimental Medicine, Hermann-Rein-Str. 3, 37075 G\"ottingen, Germany]{ Max-Planck-Institute of Experimental Medicine, G\"ottingen}  
\author{Jean Salamero}
\affiliation[Cell and Tissue Imaging Facility (PICT-IBiSA), CNRS UMR144 Institut Curie - Paris 75248, France]{CNRS UMR144 Institut Curie, Paris}
\author{Luigi Bonacina} \email{luigi.bonacina@unige.ch}  
\author{Jean-Pierre Wolf} \affiliation[GAP-Biophotonics, Universit\'e de Gen\`eve, 22 Chemin de Pinchat, 1211 Gen\`eve, Switzerland]{GAP-Biophotonics, University of Geneva}

\begin{document}

\begin{abstract}
We investigate the use of Bismuth Ferrite (BFO) nanoparticles for tumor tissue labelling in combination with infrared multi-photon excitation at 1250 nm. We report the efficient and simultaneous generation of second and third harmonic by the nanoparticles. On this basis, we set up a novel imaging protocol based on the co-localization of the two harmonic signals and demonstrate its benefits in terms of increased selectivity against endogenous background sources in tissue samples. Finally, we discuss the  use of BFO nanoparticles as mapping reference structures for correlative light-electron microscopy. 
\end{abstract}

\textit{Keywords: second harmonic generation, third harmonic generation, nanoparticles, tissue imaging, multiphoton microscopy}

% modified text begins here
The advent of multi-photon microscopy in the early nineties has revolutionized the field of optical imaging.\cite{Denk1990} This technique has proven particularly beneficial for biological studies. Nowadays, thanks to the availability of compact ultrafast sources exceeding the traditional 700-1000 nm range of Ti:Sapphire oscillators and covering  the spectral region up to 1300 nm, the way is paved for improved performances in terms of imaging penetration and  novel applications for tissue diagnostics and tumor invasion studies.\cite{Weigelin2012} The tunability of these sources allows selecting the excitation wavelength for minimizing water absorption and scattering according to sample characteristics. \cite{Andresen2009, Horton2013} Traditional nanophotonics labelling approaches (quantum dots, plasmonic nanoparticles (NPs), up-conversion NPs) display fixed optical properties often in the UV-visible spectral region and cannot fully take advantage of this spectral  extension. To circumvent wavelength limitations, a few research groups in the last years have introduced a new nanotechnological approach based on metal oxide nanocrystals with non-centrosymmetric lattice, harmonic nanoparticles (HNPs).\cite{Nakayama2007, Dempsey2012, Bonacina2013} By their crystalline structure, HNPs present very efficient nonlinear $\chi^{(2)}$ response, and can be effectively imaged using second harmonic (SH) emission as contrast mechanism efficiently responding to excitation from the UV to the mid-IR.\cite{Extermann2009} Moreover their signal is not bleaching, blinking, nor saturating because of the non-resonant character of the photo-interaction mechanism involved.\cite{LeXuan2008, Pantazis2010}   Multi-harmonic emission by HNPs has been already sparsely reported to date.\cite{Extermann2009, Cai2014, Dai2014} Extermann \textit{et al.}  firstly observed third harmonic (TH) generation in  Fe(IO$_3$)$_3$ HNPs by exciting at 1500 nm with a SH/TH intensity ratio of approximately 100.\cite{Extermann2009} 

In this work, we use 100-120 nm diameter Bismuth Ferrite (BFO) HNPs with a PEG biocompatible coating prepared for further functionalization (see   S.I. \S~1.1). BFO HNPs have been recently presented as the most promising candidates for translating HNPs to medical applications. In fact, they present a very high second order nonlinear coefficient $\langle d \rangle$=79 pm/V \cite{Schwung2014} (for comparison  $\langle d\rangle\approx$4 pm/V for BaTiO$_3$ and KNbO$_3$ HNPs)\cite{Staedler2012} and extremely good biocompatibility, in particular when PEG-coated. A thorough discussion about the effect of PEG-coated BFO HNPs on various cancer and healthy cell lines  has been recently published. \cite{Staedler2015} In this reference, the effect of BFO HNPs in terms of cell viability, membrane permeability, lysosomal mass, intracellular localization and hemolytic potential are assessed by high throughput methods at different concentrations and for several incubation times. Recently, BFO HNPs have been applied with success to novel \textit{in vitro} applications in cancer research and regenerative medicine. Apart from polymer coating for increased biocompatibility, the functionalisation of HNPs for specific targeting has been demonstrated for BaTiO$_3$.\cite{Hsieh2010a, Viskota2012, Liu2014}

%\onecolumn
\begin{figure}
\begin{center}
\includegraphics[width=10 cm]{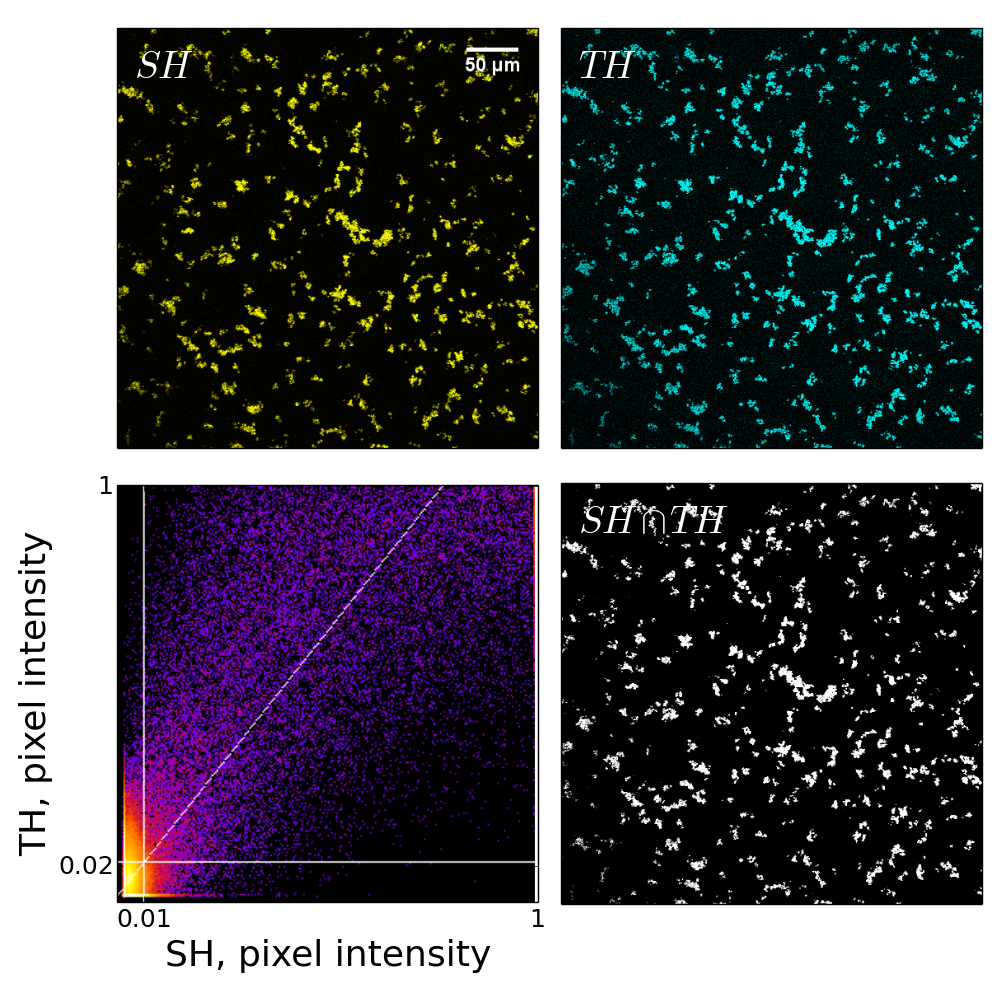}
\caption{Multi-harmonic emission by bare BFO HNPs deposited on a substrate. SH: second harmonic image. TH: third harmonic image. Scatter plot: TH \textit{vs.} SH pixel intensity. The lines parallel to the axes indicate the threshold levels determined by  Costes algorithm,\cite{Costes2004} events with intensities $>$  both thresholds are considered co-localized. The diagonal line is a linear fit to the whole dataset. SH$\cap$TH: co-localization image with co-localized pixels in white. Pixel size 0.79 $\mu$m, PSF not oversampled.}
\label{Fig:1}
\end{center}
\end{figure}
%\twocolumn

For optical nonlinear imaging we employed a Nikon A1R multiphoton upright microscope  (NIE-Nikon) coupled with an Insight Deepsee tunable laser oscillator (Spectra-Physics, 120 fs, 80 MHz, 680 - 1300 nm). With respect to standard systems based on Ti:Sapphire oscillators, the microscope was optimized for infrared transmission using tailored reflection coatings and dedicated transmission components. The nonlinear signals were epi-collected by two different long working distance  objectives (25$\times$  CFI75 APO N.A. 1.1  and 16$\times$ CFI75  N.A. 0.8) spectrally filtered by tailored pairs of dichroic mirrors and interference filters and acquired in parallel either by a standard photomultiplier (600 - 655 nm) or a GaAsP photomultiplier (385 - 492 nm).

The top-left plot in Fig.~\ref{Fig:1} shows the SH emission at 625 nm generated by BFO HNPs deposited on a substrate when excited at 1250 nm. The intensity differences among HNPs in the image reflect their size dispersion and crystal axis orientation with respect to the polarization of the excitation laser.\cite{Bonacina2007} When detecting at the TH frequency (416 nm),  HNPs appear also very bright. We have applied to this two color image the algorithm developed by Costes \textit{et al.} to detect co-localized SH/TH events.\cite{Costes2004} This recognized computational method has the advantage of automatically determining the signal thresholds for both detection channels with no user bias. Any event in the region above the two thresholds ($T_{SH}$ and $T_{TH}$) indicated by the two thick lines parallel to the axes is assumed as co-localized and the associated pixels appear white in the SH$\cap$TH plot. One can see that this region  encompasses the quasi-totality of the nanoparticles on the substrate. The diagonal line in the scatter plot is automatically determined by the algorithm as a result of a linear regression of the whole dataset. The dispersion of the data points around this reference line is not surprising: a strict linear dependence between the two normalized harmonic signals is not expected, because of their different intensity dependence ($\propto I^2$ and $\propto I^3$, respectively for SH and TH) and of other factors influencing differently the two emissions, \textit{e.g.} orientation. More importantly, a constant ratio between TH and SH by HNPs \textit{is not required} for the co-localization procedure to work. By running a test of significance, we obtain a \textit{P}-value of 1.0, indicating that all the events identified by the algorithm as co-localized are truly so from a statistical standpoint.\cite{Lifshitz1998, Schindelin2012} For a comparison, the widely used (but less informative) Pearson coefficient obtained for this data set is 0.87. Importantly, such multi-harmonic emission is not limited to BFO HNPs although for this material the ratio seems particularly favourable. In  S.I. \S~2.3,  we report the results obtained using a different HNP material (KNbO$_3$) as additional demonstration of the procedure.  

\begin{figure}[t]
\begin{center}
\includegraphics[width=12 cm]{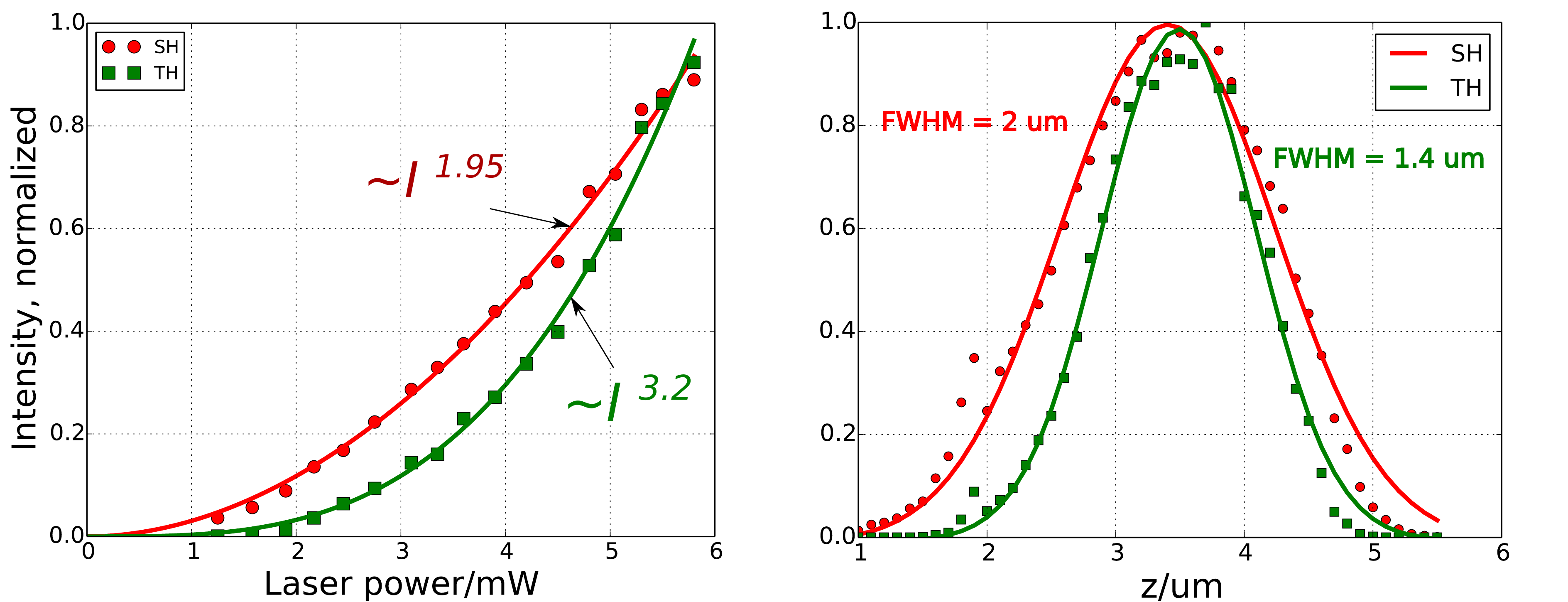}
\caption{A. SH and TH power dependence measured on an individual BFO HNP. Markers: experimental values. Thick lines: $I^n$ fits yielding $n=$1.95 and 3.2 for SH and TH, respectively. B. Nonlinear axial PSF at SH and TH obtained with a 1.1 N.A. objective and 1250 nm excitation. Markers: experimental values. Thick lines: Gaussian fits yielding FWHM=1.97 $\mu$m and 1.42 $\mu$m for SH and TH, respectively.}
\label{Fig:2}
\end{center}
\end{figure}

The normalized power dependence measured on a single BFO HNP for both SH and TH with the characteristic $I^2$ and $I^3$ dependence is reported in Fig.~\ref{Fig:2}a with the corresponding exponential fit confirming their assignment (1.95 and 3.2 for SH and TH, respectively). As expected, the SH/TH ratio changes according to excitation intensity and indeed  Dai \textit{et al.} have recently investigated this dependence in ZnO NPs for prospective  applications in display technology.\cite{Dai2014} By calibrating the spectral transmission of optics and response of the detectors, we could estimate the SH/TH BFO intensity ratio to a factor forty to one hundred at 1250 nm, depending on the intensity applied. A detailed description of the procedures used for this estimate can be found in S.I. \S~2.1. In S.I. \S~2.2 we also present a calculation on the expected emission anisotropy as a function of particle size, indicating that within the particle dimension range used in this work, no destructive interference effects are expected for the epi-detected fraction of both SH and TH emission. In Fig.~\ref{Fig:2}b, we  show how the  nonlinear axial point spread functions (PSF)  measured at the second and third order on a single sub-diffraction limited HNP are different. Because of higher nonlinearity order, TH PSF is narrower, leading to increased resolution, an aspect which might turn out to be particularly beneficial when working at long wavelengths.\cite{Zipfel2003} 

Although a large $\chi^{(3)}$ response is not surprising for large $\chi^{(2)}$ samples,\cite{Morita1993} the simultaneous collection of multiple harmonics by excitation-tunable nanometric systems  can be very advantageous for increasing selectivity in demanding applications, like ultra-sensitive detection in fluids as recently demonstrated.\cite{Geissbuehler2012} For imaging, the use of $>$ 1100 nm excitation wavelengths  ensures that standard microscope collection optics and acquisition detectors can be efficiently employed at the TH frequency. The two harmonic signals are very well spectrally separated and characterized by narrow bandwidths, which make them easily distinguishable also when using conventional fluorescence filters for detection .

To investigate the advantages of multi-harmonic detection in  relevant biomedical samples, we proceeded in imaging excised cancer tissues from a xenograft tumor model. Details of preparation are provided in  S.I. \S~1.3 and 1.4. Briefly, we analysed tumors developed in female nude mice after implantation of human breast tumor cells MDA MB 231 either subcutaneously or orthotopically in the right abdominal mammary gland fat pad. Fresh breast tumor tissue sections were obtained using a vibratome, followed either by incubation with BFO HNPs or in buffer as control.

%\onecolumn
\begin{figure}
\begin{center}
\includegraphics[width=12 cm]{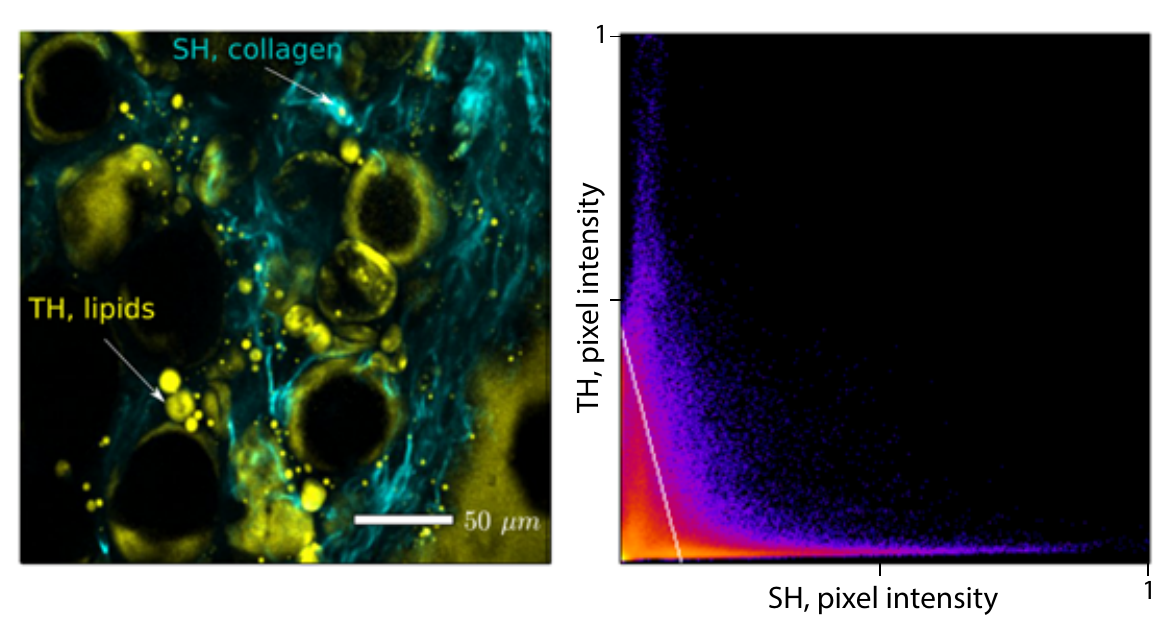}
\caption{Left. Representative multiphoton image of an unlabelled tissue section of an orthotopic breast tumor.  Cyan: SH. Yellow: TH. Right. Scatter plot TH \textit{vs.} SH. The algorithm indicates no positive events for co-localization (the diagonal line with negative slope points to  anti-colocalization;  the Pearson's coefficients is -0.09). Pixel size 0.65 $\mu$m, PSF not oversampled.}
\label{Fig:3}
\end{center}
\end{figure}
%\twocolumn

In Fig.~\ref{Fig:3} we show a multi-harmonic image of an unlabelled section from  the breast  orthotopic tumor (negative control). It is known that strong endogenous sources of second (cyan, collagen)\cite{Williams2005} and third harmonic (yellow, lipids)\cite{Debarre2005} exist in tissues. Both these tissue constituents are abundant in tumors\cite{Shoulders2009, Le2009, Kakkad2012, Weigelin2012} and are clearly present in the picture. Such harmonic background can affect the selective detection of HNPs. Some authors have already shown that HNPs can be imaged by their SH emission against collagen background (mammalian tendon).\cite{Pantazis2010, Grange2011} However, it was also shown that for individual HNPs the contrast was reduced,\cite{Grange2011} even though instrumental sensitivity was sufficient for detecting single HNPs emission.  Interestingly, the application of Costes's  algorithm fails on this image slice because of the almost total absence of  co-localization events to perform the computation. This finding is consistent with the distribution of pixels in the scatter plot,  where elements yielding simultaneously high SH and TH are sorely missing. The calculation of  Pearson's coefficient for this image yields a negative value (-0.09).

\begin{figure}

\includegraphics[width=16 cm]{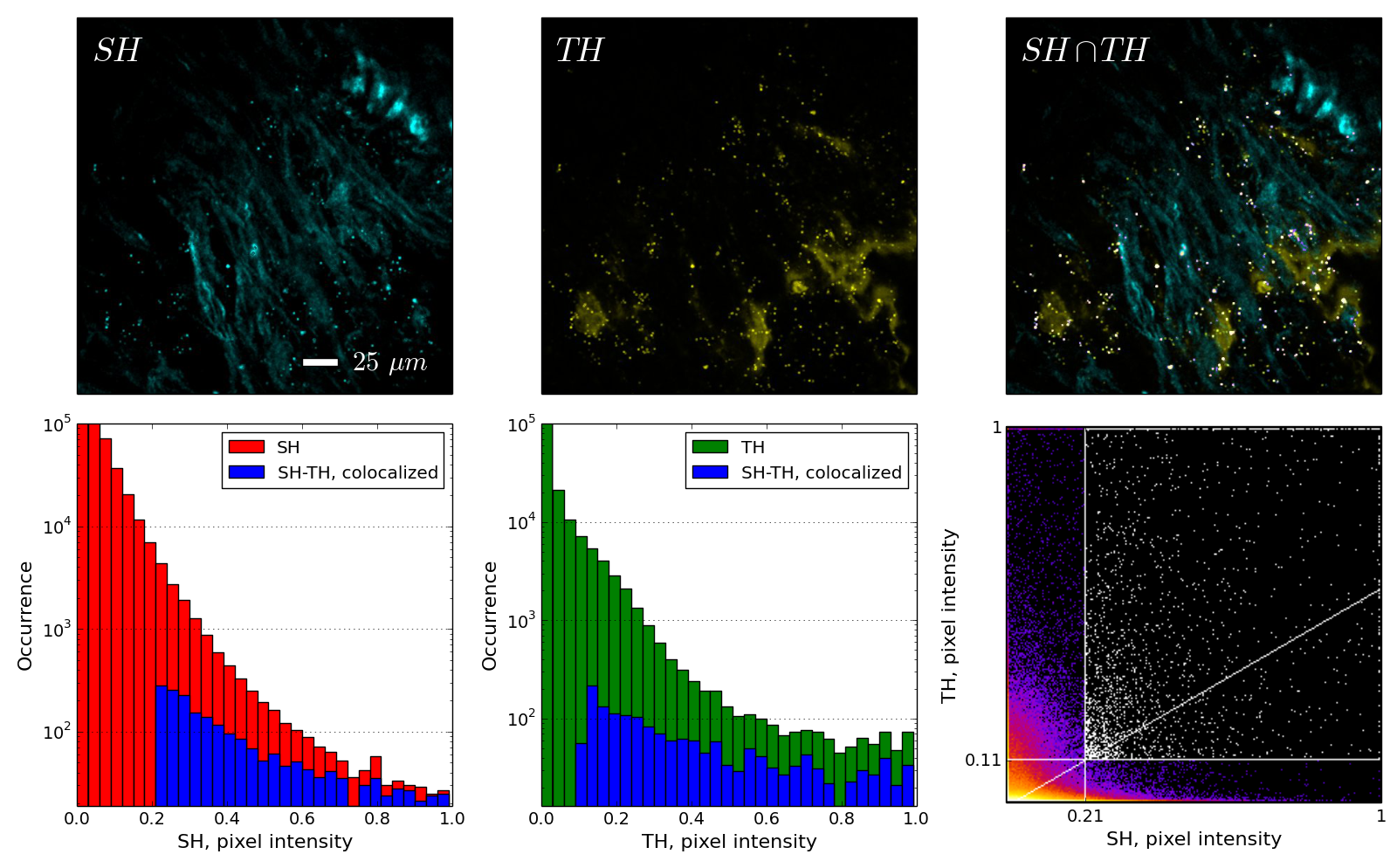}
\caption{Representative multiphoton images of subcutaneous  breast tumor tissue labelled with BFO HNPs. First row. SH (cyan), TH (yellow), and  SH$\cap$TH merged image with  co-localized pixels in white. Second row. Semi-log SH and TH intensity histograms. Blue bars correspond to co-localized events.  Scatter plot TH \textit{vs.} SH. As in Fig~\ref{Fig:1}, the lines parallel to the axes indicate the automatically determined threshold values and the diagonal line a linear fit to the whole dataset. Only the events above the thresholds (white pixels) are assumed as co-localized. Pixel size 0.65 $\mu$m, PSF not oversampled.}
\label{Fig:4}

\end{figure}

The first column of Fig.~\ref{Fig:4}, displays the SH signal of a tissue section from the subcutaneous  breast tumor which was incubated with BFO HNPs. Collagen structures are clearly evident as diagonal stripes, the presence of small bright spots (sometimes at the limit of pixel resolution) is more pronounced than in the negative control of Fig.~\ref{Fig:3} and points to the presence of HNPs in the sample. Likewise, TH image in the central row shows the presence of small spots, together with other larger structures with different morphology with respect to collagen, which can be ascribed to lipids.\cite{Le2009, Weigelin2012} The results of the co-localization procedure (SH$\cap$TH, rightmost plot) enables highlighting in white exclusively pixels which show a simultaneous multi-harmonic emission and that, on the basis of the findings discussed in relation to Figs.~\ref{Fig:1} and \ref{Fig:3}, can be safely associated with the presence of BFO HNPs. The second row of Fig.~\ref{Fig:4} provides a more quantitative analysis of this co-localization effect. The semi-log histograms of pixel intensity both at the SH and TH covers the whole detection dynamic range. Co-localization bars (in blue) indicate the number of occurrences which can be simultaneously attributed to both harmonic channels. One can see that the discrimination of HNPs signal against background emission of endogenous sources cannot be based simply on intensity, as in both detection channels a relevant fraction of events at high intensity are not co-localizing. When comparing the intensity emitted by endogenous structures with that of HNPs, one has to take into account that the former  despite their lower nonlinear efficiency (e.g. $<d>=0.94$ pm/V for collagen\cite{Stoller2003}) are generally characterized by much larger dimensions (fulfilling or exceeding the focal volume) therefore the squared-volume dependence of harmonic signals can easily counteract the harmonic generation efficiency difference.

We cannot exclude that a few low intensity multi-harmonic positive events are not accounted for by the algorithm as inevitable in the case of automatic image analysis.\cite{Costes2004}  On the other hand, false positive detections seem unlikely. Although we used relatively broadband fluorescence filters as compared to the narrow SH and TH emission bandwidths, we carefully ensured that no autofluorescence is emitted upon two- and three-photon absorption when working at 1200 nm with the excitation settings employed here (for a complete analysis refer to S.I. \S~3). Clearly, a preliminary assessment of the anti co-localization of endogenous sources in a negative control sample (Fig.~\ref{Fig:3} in the present case) is mandatory for the success of the approach.

%\end{center}

The application of multi-harmonic correlation for increasing selectivity by background rejection in optically congested environment like tissues can be performed in real-time, and therefore opens the way to HNPs tracking protocols\cite{Magouroux2012, Macias2014} with high selectivity and minimal image processing requirements. Moreover, autofluorescence can be minimized when working in this wavelength range, avoiding endogenous fluorophores like riboflavins and NADH which display two-photon absorption bands peaked at wavelengths $<$1000 nm.\cite{Zipfel2003} In addition, it is worth noting that resolution is generally not an issue for detecting HNPs. As demonstrated by the Beaurepaire group, it is advantageous when imaging thick samples by multiphoton excitation to employ objectives with  low-magnification and large field of view / N.A.  to epi-collect efficiently multiple scattered photons.\cite{Debarre2007}

\begin{figure}

\includegraphics[width=12 cm]{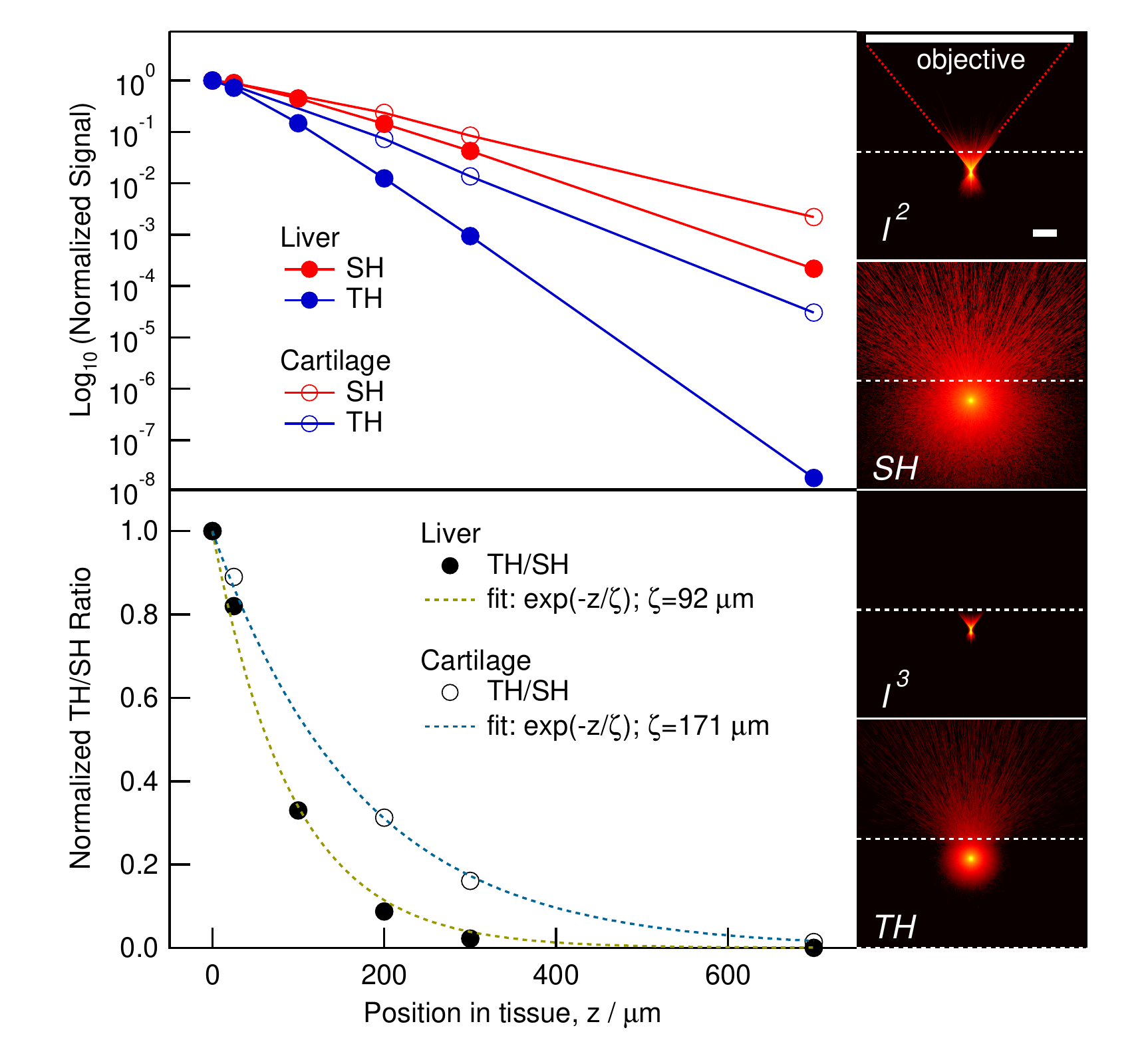}
\caption{Results of Monte Carlo light transport simulations. Upper plot. Normalized detected SH and TH signals (semi-log scale) as a function of HNP position $z$ for two relevant tissue types: liver and cartilage. Lower plot. Normalized TH/SH  ratio calculated from the same dataset. Right panels. Two dimensional slices of the simulation volume in the case of a HNP embedded at $z=300~\mu$m in liver, reporting from top to bottom: the excitation intensity squared, the SH emission pattern, the excitation intensity cubed, and the TH emission pattern. The upper side of each image correspond to the objective aperture, the dashed line indicates the position of tissue surface. The intensity logarithmic colormaps are  conserved  between focal intensity plots (spanning 5 orders of magnitude) and TH, SH emission plots (seven orders of magnitude) allowing a direct comparison.}
\label{Fig:5}

\end{figure}

The measurements presented were performed with peak pulse intensity at the sample  going from 300 to 650 GW/cm$^2$. Even by assuming the highest of these two values (calculated in the femtosecond regime and corresponding to 16 mW average laser power) and considering the scanning speed used for raster imaging, which corresponds to a pulse density of 1.2 $\times 10 ^3~\mu$m$^{-2}$,\cite{Supatto2005} we can calculate that these conditions are suitable for imaging of biological samples. Indeed these figures are sensibly lower than the photo-induced tissue damage threshold established by Supatto \textit{et al.} on living drosophila embryos.\cite{Supatto2005} As a further assessment, the highly cited paper by K\"oning on cellular response to near infrared femtosecond excitation, sets a damage threshold on cell viability and reproduction at 1 TW/cm$^2$.\cite{Konig1997}  Although we approach this intensity value in absolute terms, it should be noted that this estimate was derived at a shorter wavelength (730-800 nm instead of $>$1200 nm) implying a lower order multiphoton process for absorption by endogenous molecules and at a much longer dwell time (80 $\mu$s as compared to 4 $\mu$s here).

To further assess the actual penetration limits of the proposed HNP-based multi-harmonic imaging protocol, we have simulated the optical response of a HNP embedded at depth $z$ in a tissue and calculated the theoretical epi-detected signals.   This numerical analysis aims at determining how the TH/SH ratio obtained at the uppermost surface of a three dimensional $z$-stack is modified upon penetration in the tissue. Simulations are based on a free-ware $C$-code for light propagation\cite{MCxyzwebpage, Wang1995} that we modified to account for nonlinear excitation. We assume that a 1200 nm input beam is focused by a N.A. 0.8, 3 mm WD objective in two different kinds of tissue slabs (rat liver and ear cartilage) characterized by the sets of optical properties ($\mu_a(\lambda)$, $\mu_s(\lambda)$, $g(\lambda)$) reported in S.I. \S~5.\cite{BoasHandBook} The simulation volume is composed by 700$^3$ voxels of 7 $\mu$m side. The HNP placed at the focus is assumed to irradiate isotropically at the SH and TH with an efficiency depending respectively on the local $I^2$ and $I^3$ excitation intensity. We then calculate the fraction of SH and TH photons emitted by the HNP reaching the objective and from these value, the  TH/SH value. This ratio is assumed to be 1 at the entrance surface. In the lower plot of Fig.~\ref{Fig:5}, one can observe an exponential decrease of the TH/SH ratio as a function of depth $z$. The observed decrease is more prominent for  liver  than for  ear cartilage, consistently with the higher scattering and absorption at the TH wavelength for the former tissue. This finding can be appreciated in the series of images on the right of Fig. \ref{Fig:5}, reporting the normalized focal intensity and emission patterns for a HNP embedded at 300 $\mu$m depth in a liver tissue. As sketched in the top panel, the upper side of each image corresponds to the objective lens and the red dotted lines the focusing beam. The horizontal dashed line defines the position of tissue surface. Comparing the SH and TH emission patterns, one can see how the penetration of TH in the tissue is strongly limited by absorption as compared to SH. As a general trend, we can deduce that for these two representative tissues, the initial TH/SH ratio drops to its $1/e$ value within the 100-200 $\mu$m range. Based on the  fact that the co-localization algorithm does not require a fixed ratio and that we could roughly co-localize events with TH/SH  variations within one order of magnitude (see scatter plot in Fig.~\ref{Fig:1}), for a given set of detection parameters, we can extrapolate a working range of approximately 300 $\mu$m in a cartilage-kind of tissue and approximately 200 $\mu$m in the optically more extreme liver case. These values are compatible with the requirements of many cell tracking applications.          

\begin{center}
\begin{figure}
\includegraphics[width=8 cm]{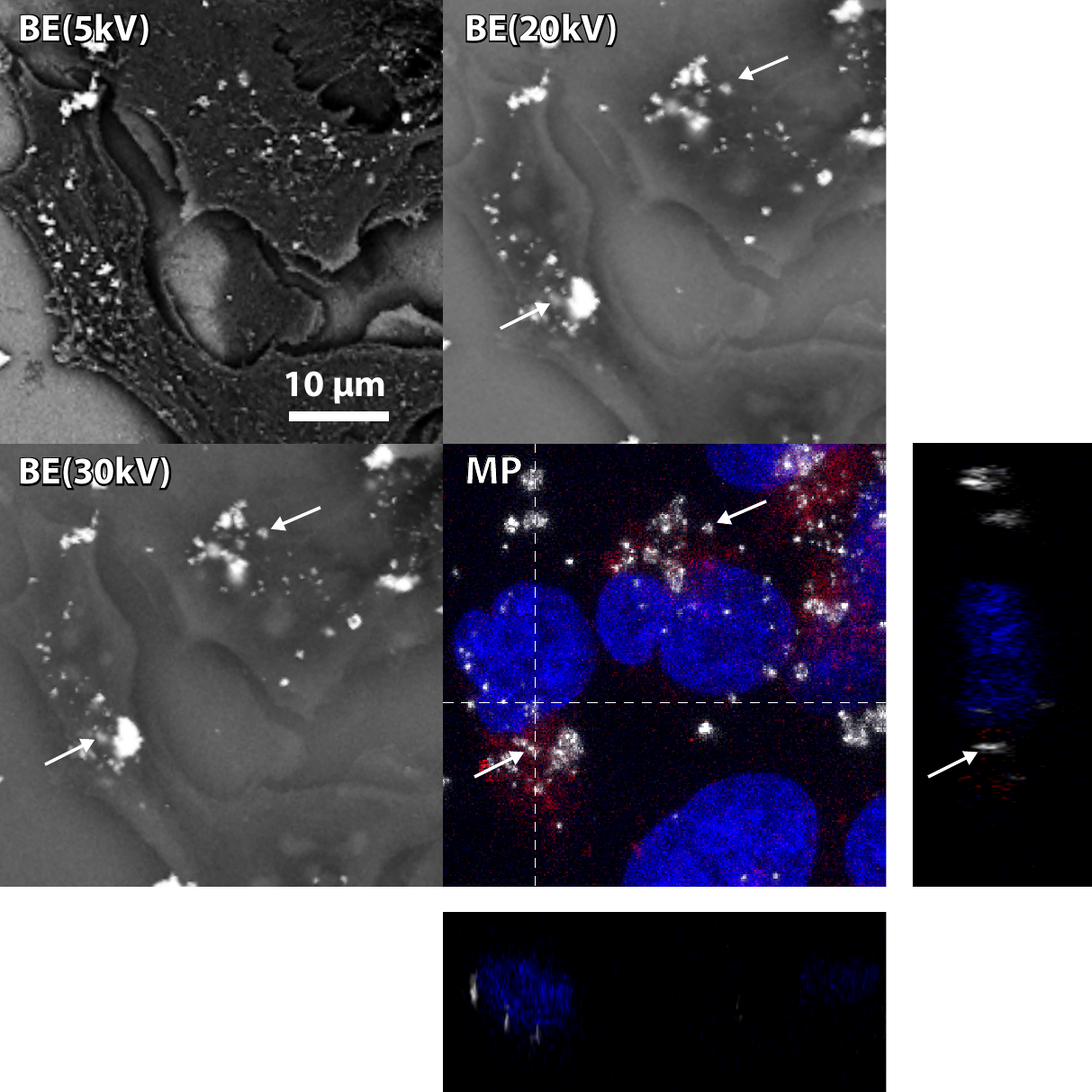}
\caption{Demonstration of CLEM imaging of  MDA MB 231 cells labelled by BFO HNPs.  BE: secondary electrons SEM images at 5, 20  and 30 kV. MP: Multiphoton maximal intensity projection image with two vertical cuts corresponding to the dashed lines. Red: Nile red fluorescence staining lipids (membranes). Blue: DAPI fluorescence (nuclei). White:   SH by HNPs. Arrows indicate throughout the different images two exemplary clusters appearing only at high SEM voltages (20 and 30 kV).}
\label{Fig:6}
\end{figure}
\end{center}

In the light of previous results, an interesting development for the use of HNPs for tissue imaging is their application as reference mapping structures for correlating light (multi-photon) and electron microscopy (CLEM), facilitating the retrieval of specific regions of interest going from one technique to the other.\cite{Glenn2012} In this respect, BFO HNPs are particularly appealing, as they are electron dense and are expected to provide good contrast in electron-based imaging like TEM and SEM. Figure \ref{Fig:6} shows a SEM representative image, including several cells imaged at different voltages in backscattered electron (BE) mode. BE  ($>$50 eV) are preferable in this case, as they are known to be sensitive to the composition of the specimen and display brighter contrast for heavier elements.  Overall, one can notice the good contrast provided by SEM for BFO HNPs. By changing accelerating voltage one can modify the scattering region of the incident electron beam (S.I. \S~6). While the 5 kV image reports details of cell's surface and indicates the presence of individual particles or small HNPs clusters, when SEM voltage is increased to 20 and 30 kV, larger aggregates located below the cell surface appear as indicated by the arrows. The presence and position of these aggregates correlate well with previous studies indicating that BFO HNPs initially  adhere to the cell membrane (2 h) and are successively internalized with increasing exposure time and  tend to concentrate in intracellular organelles, such as endosomes or lysosomes.\cite{Staedler2015} Finally, panel MP shows a maximal intensity projection image of the same R.O.I. imaged by multi-photon microscopy.

In this case, fluorescent labels were used to stain nuclei (DAPI, blue) and lipids such as cell membranes (Nile Red, red). SH from HNPs appears in white. One can see that HNPs tend to localize at the membranes, in particular in region with higher density subcellular membrane compartments where they are thicker (red intensity higher). Two vertical projections corresponding to the sections indicated by the dashed lines are also reported. Within the optical resolution, they indicate that HNPs are mainly located inside the membranes or at cell upper surfaces confirming an already observed tendency to avoid nuclear region and colocalize in lysosomes or at cell surface.\cite{Staedler2015}

In conclusion, we have shown that multi-harmonic emission of BFO HNPs can be easily detected by  multi-photon microscopy when using excitation $>$1100 nm. Very advantageously, the intensity difference among second and third nonlinear response is not large. Using the right combination of detectors for the different spectral ranges (GaAsP and standard photomultipliers) the two  signals can be acquired simultaneously using standard settings. Based on this result, we have  demonstrated that the co-localization of SH and TH allow identifying with high selectivity HNPs in a complex optical environment presenting endogenous sources of fluorescence and harmonic generation, an excised xenograft tumor tissue in our experiment.  The image processing necessary for this approach relies on simple two-channel co-localization, allowing its  real-time use in demanding imaging applications. Numerical simulations of light transport indicate that the proposed imaging protocol can be performed up to a thickness of a few hundreds micrometers, which are relevant length scales for cell tracking applications. Moreover, by additional electron microscopy measurements, we have shown that BFO HNPs could prospectively serve as localization fiduciaries  in advanced CLEM studies.\cite{Goetz2011}
% in vivo deleted

%\begin{figure}
%\begin{center}
%\includegraphics[width=8 cm]{negative.pdf}
%\caption{bla bla}
%\label{Fig:1}
%\end{center}
%\end{figure}

%\begin{figure}
%\begin{center}
%\includegraphics[width=8 cm]{tissue.pdf}
%\caption{bla bla}
%\label{Fig:1}
%\end{center}
%\end{figure}

\begin{acknowledgement}
This research was partially supported by the European FP7 Research Project NAMDIATREAM (NMP4-LA-2010-
246479, http://www.namdiatream.eu), Fondation pour la Recherche M\'edicale ( FRM n. DGE20111123020),  and the Canc\'erop\^ole Ile de France (n. 2012-2-EML-04-IC-1). 246479, http://www.namdiatream.eu). The financial support by the  NCCR Molecular  Ultrafast  Science  and  Technology  (NCCR MUST), a research instrument of the  Swiss National Science Foundation (SNSF), is also  acknowledged. The study was performed in the context of the European COST Action MP1302 Nanospectroscopy. GAP-Biophotonics authors are grateful to Michel Moret for technical support.

\end{acknowledgement}
%\begin{suppinfo}
%Additional details on the experimental procedures.
%\end{suppinfo}

%
%\bibliographystyle{unsrt}
%\bibliography{HNPs}

\providecommand*\mcitethebibliography{\thebibliography}
\csname @ifundefined\endcsname{endmcitethebibliography}
  {\let\endmcitethebibliography\endthebibliography}{}

%\newpage
%\textbf{For Table of Contents Use Only}
%
%
%Title: \textit{Simultaneous Multi-Harmonic Imaging of Nanoparticles in Tissues for Increased Selectivity}
%
%
%Authors: Andrii Rogov, Marie Irondelle,Fernanda Ramos-Gomes, Julia Bode, Davide Staedler, Solene Passemard, Sebastien Courvoisier, Yasuaki Yamamoto,François Waharte,Daniel Ciepielewski,Philippe Rideau, Sandrine Gerber-Lemaire, Frauke Alves, Jean Salamero, Luigi Bonacina, and Jean-Pierre Wolf
%
%\begin{center}
%\begin{figure}
%\includegraphics[width=8 cm]{TOC.png}
%\caption*{We present a protocol based on the simultaneous detection of multi-order nonlinear response by Harmonic Nanoparticles to enhance imaging specificity in tissues. }
%\label{Fig:TOC}
%\end{figure}
%\end{center}
%

\end{document}